\documentclass[twocolumn,prl,superscriptaddress,bibnotes,notitlepage,nofootinbib]{revtex4-1}
\usepackage{color}
\definecolor{red}{rgb}{0.75,0,0}
\definecolor{blue}{rgb}{0,0,0.75}
\definecolor{green}{rgb}{0,0.5,0}

\usepackage{amsmath}
\usepackage{amssymb}
\usepackage{lipsum}
\usepackage{bm}
\usepackage{xcolor}
\usepackage{graphicx}
\usepackage{verbatim}
\usepackage[T1]{fontenc}
\usepackage{hyperref}

\def\be{\begin{equation}}
\def\ee{\end{equation}}
\def\bea{\begin{eqnarray}}
\def\eea{\end{eqnarray}}

\def\besub{\begin{subequations}}
\def\eesub{\end{subequations}}

\def\bwd{\begin{widetext}}
\def\ewd{\end{widetext}}

\definecolor{ao(english)}{rgb}{0.0, 0.5, 0.0}
\definecolor{armygreen}{rgb}{0.29, 0.33, 0.13}
\definecolor{auburn}{rgb}{0.43, 0.21, 0.1}
\definecolor{brightmaroon}{rgb}{0.76, 0.13, 0.28}
\definecolor{cadmiumred}{rgb}{0.89, 0.0, 0.13}
\definecolor{carnelian}{rgb}{0.7, 0.11, 0.11}
\definecolor{cornellred}{rgb}{0.7, 0.11, 0.11}
\definecolor{crimsonglory}{rgb}{0.75, 0.0, 0.2}
\definecolor{orangeyellow}{rgb}{0.3, 0.2, 0.2}
\definecolor{fluorescentorange}{rgb}{1.0, 0.75, 0.0}
\definecolor{gamboge}{rgb}{0.89, 0.61, 0.06}
\newcommand{\bsf}[1]{\textsf{\textbf{#1}}}

\newcommand{\AMR}[1]{\textcolor{black}{#1}}
\newcommand{\AMRR}[1]{\textcolor{black}{#1}}

\begin{document}
\title{\AMR{Stable long-range uniaxial order in active fluids at two-dimensional interfaces}
}
\author{Ananyo Maitra}
\email{nyomaitra07@gmail.com}
\affiliation{Sorbonne Universit\'{e} and CNRS, Laboratoire Jean Perrin, F-75005, Paris, France}

\begin{abstract}
\AMR{
	I show that two-dimensional nematic order in an active fluid film can be stable and long-ranged if exchange of particles and momentum with an ambient three-dimensional fluid is allowed. Number-conserving films present an instability, with \AMRR{a fastest-growing mode} independent of activity at high activities or low concentrations. Motility can stabilize long-range polar order on such immersed surfaces even when the swimmer number in the film is conserved. Active ordering in immersed surfaces is much richer than previously expected, with functional consequences for active transport, and suggests that a broader exploration of parameters in current and future experiments would display the new features predicted here.}	

\end{abstract}

\maketitle

Active matter is driven out of equilibrium by continuous supply of energy at microscopic scales, leading to macroscopic forces and currents \cite{SRJSTAT, LPDJSTAT, RMP, Curie1, Curie2, Prost_nat, Sal1, SRrev, CatTal1, Chate1, TonTuRam, Toner_Tu}. 
These active forces have spectacular consequence on the phase behaviour of ordered states, one of the most notable of which is the threshold-free instability of uniaxial phases \AMR{-- both nematic and polar --} in incompressible, bulk fluids \cite{Aditi1, Voit1, RMP} first noted in a seminal paper by Simha and Ramaswamy \cite{Aditi1}. \AMR{This has led to the expectation that uniaxial active phases cannot be realised in any momentum-conserved fluid and, in particular, nematic or polar phases at interfaces in bulk fluids \cite{Dogic1, Sagues} must generically devolve into spatiotemporally chaotic states with a characteristic wavelength that decreases with activity \cite{Guo1, Guo2, Sagues}.}
\begin{figure}
  \centering
  \includegraphics[width=8cm]{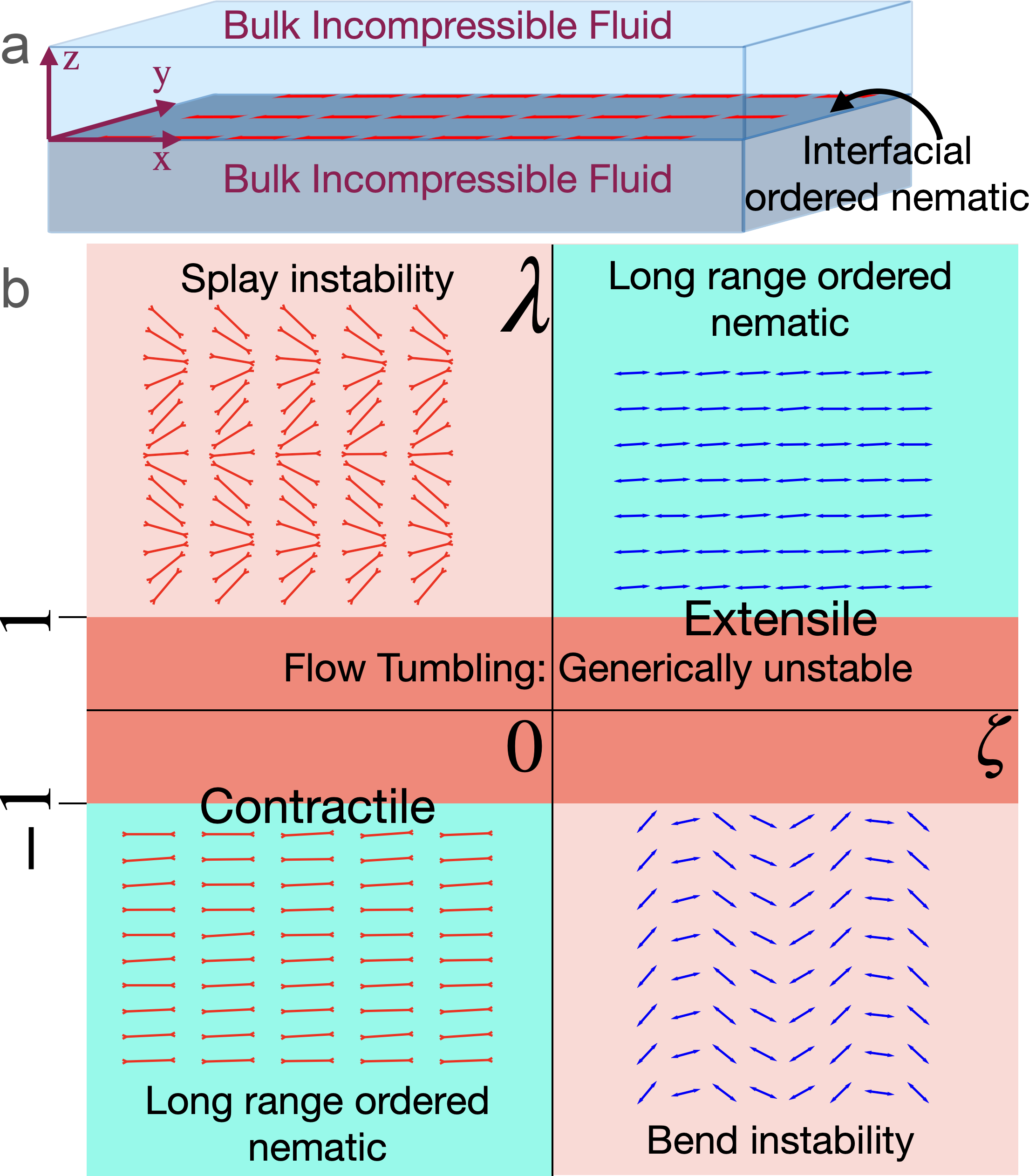}
\caption{\AMR{a. Experimental geometry for the predicted two-dimensional, long-range ordered nematic phase. The active nematic units order at the fluid-fluid interface at $z=0$. b. Stability diagram of active nematics in the $z=0$ interfacial plane. $\zeta$ is the coefficient of active stress and $\lambda$ is the flow alignment parameter. For $|\lambda|>1$ and $\zeta\lambda>0$, a long-range order nematic phase is realised. Either when $\lambda\zeta<0$ or for $|\lambda|<1$ the ordered phase is destabilised for splay in contractile ($\zeta<0$) systems or for bend perturbations in extensile ($\zeta>0$) systems.}}
\label{Fig_summ}
\end{figure}

\AMR{In this paper, I consider active uniaxial ordering at immersed interfaces \cite{Guo1, Guo2} and membranes in bulk, momentum-conserved, incompressible fluids at scales larger than the Saffman-Delbr\"{u}ck length \cite{Saffman} and show that this is only correct in a particular limit and hides a much richer dynamics. Importantly, I show that \emph{long-range ordered} nematic (Fig. \ref{Fig_summ}) and polar flocks should be observable in this system despite bulk momentum conservation. This is a direct consequence of the fact that {interfaces} and membranes in incompressible fluids are \emph{not} generically incompressible in two dimensions \cite{com1, com2, com3}; in fact, they must be compressible at large scales due to fluctuations \cite{Cai}.
}

\AMR{In biological systems, uniaxial ordering is more likely to be associated with interfaces or membranes \cite{Ano_mem}, such as those enclosing cells, than in the bulk and may be relevant for active transport \cite{Mad}. \AMRR{The two-dimensional flow in this case is \emph{compressible} as argued in \cite{Sal1, Salbreaux}.} Swimmers, such as bacteria \cite{bac_int1, bac_int2, bac_int3, bac_int4}, generally aggregate at interfaces and, if they are elongated, can form effectively two-dimensional uniaxial phases. 
	Interfacial order is also relevant for a widely used experimental geometry for studying pattern formation in active nematics \cite{Dogic1, Sagues, Ano_und, Ramin_und} composed of a thin layer of motor-microtubule filaments in a three-dimensional bulk fluid \cite{Footnote1}. 
}



\AMR{The key results of this paper, \AMRR{summarised in the table below}, are i. a two-dimensional \emph{long-range ordered} (LRO) active nematic state can be realised at an interface between bulk, momentum-conserved fluids if the number of nematogens \emph{at the interface} is not conserved (the total number of particles in the system is conserved) yielding a mechanism for taming the Simha-Ramaswamy instability (Fig. \ref{Fig_summ}). This is the first prediction of a two-dimensional LRO active nematic phase and its hydrodynamic properties belong in a new universality class which I characterise. ii. Homogeneous nematic order is \emph{generically} destroyed for a species living \emph{at an interface or membrane}. 
	The form of the instability, for a highly compressible interfacial species (low concentration) is however distinct from that of an essentially incompressible one with the \AMRR{growth-rate} becoming independent of activity.
	iii. An LRO polar phase is stabilised by large motilities even for a conserved interfacial species. In this case, the concentration fluctuations violate the law of large numbers with the R.M.S. number fluctuations in a region containing on average  $\langle N\rangle$ swimmers scaling as $\langle N\rangle^{3/4}$ instead of $\sqrt{\langle N\rangle}$ as it would in all equilibrium systems.
}
I now discuss these results in detail.

\begin{table}
	\begin{ruledtabular}
		\begin{tabular}{p{0.45in} p{0.9in}p{0.05in}p{0.9in}p{0.05in}p{0.9in}}
			& non-conserved concentration & & Highly compressible & & Incompressible\\
			\hline
			Nematic & LRO ($\zeta\lambda>1$ and $|\lambda|>1$) & & Unstable; growth-rate $\zeta$-independent  & & Unstable \\
			
			Polar & LRO ($\zeta\lambda>1$ and $|\lambda|>1$) & & LRO  (large moitlity); GNF: RMS fluctuations $\sim\langle N\rangle^{3/4}$ & & Unstable \\
		\end{tabular}
	\end{ruledtabular}
\end{table}

I consider
an active nematic phase at a perfectly flat, thin interfacial layer or membrane at $z=0$ \AMR{immersed in a bulk fluid}. \AMR{I first consider the case in which the number of active nematogens \emph{at the interface} is not conserved i.e., the orientable particles, \AMRR{or the monomers composing them,} can diffuse in and out of the interface. \AMRR{A particularly simple realisation of this, relevant for cellular cortex or motor-microtubule films, is when monomers are dispersed homogeneously in three-dimensions, but they can associate (and dissociate) to form nematogenic units at the interface. In this case, the monomer concentration (including free monomers and those in the nematogens) has a linearised dynamics $\dot{\rho}=D_3\nabla^2\rho$ in three-dimensions. The concentration of nematogens $\rho_N$ at the interface is not conserved and has a mean value $\langle\rho_N\rangle=k_0\langle\rho(z=0)\rangle/k_1$ where $k_0$ and $k_1$ are the association and dissociation rates respectively.} \cite{supp}.}
	For simplicity, I take the fluids \AMR{above and below the interface} to be viscosity-matched. The three-dimensional bulk fluid velocity ${\bf V}({\bf r}, t)\equiv{\bf V}({\bf r}_\perp, z, t)$, where ${\bf r}_\perp\equiv(x,y)$, is incompressible i.e., $\nabla\cdot{\bf V}=0$. The two-dimensional nematic order is taken to be along $\hat{x}$ and $\theta$ is taken to be the deviation of the local nematic order from $\hat{x}$.
Defining ${\bf v}({\bf r}_\perp, t)={\bf V}_\perp({\bf r}_\perp, z=0, t)$, the linearised equation of motion for small angular fluctuations \AMR{about the ordered phase is}
\begin{equation}
\label{angeq}
\partial_t\theta=\Omega_{xy}-\lambda A_{xy}-\Gamma_\theta\frac{\delta F}{\delta\theta}+\xi,
\end{equation}
where $\boldsymbol{\Omega}=(1/2)[\nabla_\perp{\bf v}-(\nabla_\perp{\bf v})^T]$ is the planar vorticity tensor at the interface and ${\bsf A}=(1/2)[\nabla_\perp{\bf v}+(\nabla_\perp{\bf v})^T]$ is the interfacial strain-rate tensor, $F=\int d{\bf r}_\perp (K/2)(\nabla_\perp\theta)^2$ and $\xi({\bf r}_\perp, t)$ is a Gaussian, white noise with the correlator $\langle\xi({\bf r}_\perp,t)\xi({\bf r}_\perp',t')\rangle=2\Delta\delta({\bf r}_\perp-{\bf r}_\perp')\delta(t-t')$. The linearised equation for the three-dimensional velocity field is 
\begin{equation}
\label{vel3d}
\eta\nabla^2{\bf V}=\nabla\Pi+\zeta(\partial_y\theta\hat{x}+\partial_x\theta\hat{y})\delta(z)+{\bf f}^{p}\delta(z)+\boldsymbol{\xi}^v,
\end{equation}
where $\Pi$ is the three-dimensional pressure \AMRR{(in which three-dimensional monomeric fluctuations have been absorbed)} enforcing the three-dimensional incompressibility constraint $\nabla\cdot{\bf V}=0$, \AMR{$\zeta$ is the standard linearised active force density  \cite{Aditi1, RMP, SRrev} with $\zeta>0$ and $\zeta<0$ denoting extensile and contractile forces respectively \cite{supp},}
and $\boldsymbol{\xi}^v$ is a conserving \AMR{white} noise with $\langle\boldsymbol{\xi}^v({\bf r},t)\boldsymbol{\xi}^v({\bf r}',t')\rangle=-2\Delta^v{\bsf I}\nabla^2\delta({\bf r}-{\bf r}')\delta(t-t')$. The linearised passive force density ${\bf f}^p=[(1+\lambda)/2]\partial_y(\delta F/\delta\theta)\hat{x}-[(1-\lambda)/2]\partial_x(\delta F/\delta\theta)\hat{y}$ \AMR{appears with two more gradients than the active force density and therefore doesn't affect the long-wavelength, low-frequency dynamics \cite{supp}.} 
Since nematogens must lie parallel to the interface, there is no component of the active and passive nematic forces in the $\hat{z}$ direction and they force the fluid only at $z=0$. \AMR{Therefore, solving \eqref{vel3d},}
\begin{equation}
\label{velavg}
{\bf v}=\int ^\infty_{-\infty}\frac{dq_z}{2\pi}{\bf V}_\perp=-\frac{i\zeta(q_y^3\hat{x}+q_x^3\hat{y})}{2\eta|q_\perp|^3}\theta+\bar{\xi}^v_x\hat{x}+\bar{\xi}^v_y\hat{y}
\end{equation}
where ${\bf q}_\perp\equiv q_x\hat{x}+q_y\hat{y}$ and $\bar{\xi}^v_x$ and $\bar{\xi}^v_y$ are the noises arising from $\boldsymbol{\xi}^v$ whose correlations are displayed in \cite{supp}. 
Plugging \eqref{velavg} in \eqref{angeq}, and defining $\phi$ as the angle between ${\bf q}_\perp$ and $\hat{x}$, I obtain, to leading order in wavenumbers,
\begin{equation}
\label{angrelang}
\partial_t\theta=\frac{\zeta|q_\perp|}{4\eta}\left[\cos(2\phi)[1-\lambda\cos(2\phi)]-\frac{\lambda}{2}\sin^2(2\phi)\right]+\bar{\xi}+\xi.
\end{equation}
where $\bar{\xi}$ is the noise from the velocity field. I show in \cite{supp} that $\langle\bar{\xi}({\bf q}_\perp,t)\bar{\xi}({\bf q}'_\perp,t)\rangle$ \AMR{vanishes as} $\sim q_\perp$ and is therefore subdominant to $\xi$ \AMR{at small wavenumbers}. \AMR{Eq. \eqref{angrelang} implies that the relaxation rate for angular fluctuations is \emph{positive} for all $\phi$ when $|\lambda|>1$ -- i.e., flow-aligning -- and $\zeta\lambda>0$ \cite{supp}. This implies the existence of a \emph{stable} nematic phase even in this bulk momentum conserved system. The ordered phase is unstable to fluctuations in flow-tumbling systems $|\lambda|<1$ for \emph{either} sign of activity -- extensile systems ($\zeta>0$) are unstable for $\phi\approx 0$, i.e., bend, and contractile systems ($\zeta<0$) are unstable for splay, $\phi\approx\pi/2$ -- or even in flow-aligning systems when $\zeta\lambda<0$. In these cases, as in \cite{Guo1, Guo2, Sagues}, a patterned state without nematic order is predicted with a wavenumber, calculated by retaining the $-\Gamma_\theta K q_\perp^2$ term  from \eqref{angeq} in \eqref{angrelang}, $q_\perp^c=\text{Max}[\zeta\{2\cos2\phi-{\lambda}(1+\cos^22\phi)\}]/(8\eta\Gamma_\theta K)$, where $\text{Max}$ denotes the maximum value of a function.}

\AMR{The unusual stability of the nematic phase when $\zeta\lambda>0$ and $|\lambda|>1$ is a direct consequence of the lack of two dimensional incompressibility of the layer. It arises due to the term $\propto \sin^2(2\phi)$ in \eqref{angrelang} which would have been missing if the layer were considered to be incompressible in two dimensions \cite{Guo1, Guo2, Sagues} since from \eqref{velavg}, $i{\bf q}_\perp\cdot{\bf v}=(\zeta\theta/2\eta)(q_xq_y/|q_\perp|)=(\zeta\theta/4\eta)|q_\perp|\sin 2\phi$. Since layer incompressibility is an extra constraint which is not expected to be valid in general, ordered nematic phases associated with interfaces should be observable even in momentum conserved biological and biomimetic systems.
}

\AMR{I now demonstrate that the interface-associated nematic phase has LRO.} \AMR{From} Eq. \eqref{angrelang}, the static structure factor of angular fluctuations when $\zeta\lambda>0$ and $|\lambda|>1$ \AMR{is}
\begin{equation}
	\label{ssfang}
\langle|\theta({\bf q}_\perp,t)|^2\rangle=\frac{8\eta\Delta}{\zeta|q_\perp|[-2\cos(2\phi)+\lambda\{1+\cos^2(2\phi)\}]}.
\end{equation}
This decays as $\sim 1/|q_\perp|$ along \emph{all} directions of the wavevector space unlike the static structure factor of passive nematics or active nematics on substrates both of which decay as $\sim 1/q_\perp^2$ \cite{Aditi2, Ano_apol, RMP}. \AMR{Therefore,  $\langle\theta({\bf r}_\perp, t)^2\rangle=\int (d{\bf q}_\perp/4\pi^2)\langle|\theta({\bf q}_\perp,t)|^2\rangle$ doesn't diverge with system size implying that for sufficiently small $\Delta$, the angular fluctuations \AMR{are} small enough to preserve LRO.}
Furthermore,
$\langle\theta(0, t)\theta({\bf r}_\perp, t)\rangle=\int (d{\bf q}_\perp/4\pi^2)e^{i{\bf q}_\perp\cdot{\bf r}_\perp}\langle|\theta({\bf q}_\perp,t)|^2\rangle\sim 1/|{\bf r}_\perp|$. Since the fluctuations decay at large scales as $1/|{\bf r}_\perp|$, the behaviour of the angle field under the scalings $x\to b x$, $y\to b^\mu y$, $t\to b^z t$ and $\theta\to b^\chi\theta$, with $\mu$ being the anisotropy exponent, $z$ being the dynamical exponent and $\chi$ being the roughness exponent, directly yields $\chi=-1/2$ within the linear theory \AMR{whose negativity is the signature of} LRO. \AMR{The nematic LRO phase is promoted by activity and fluid dynamics-induced long-range interactions whose spatial character is analogous to dipolar interactions which induce LRO in passive, two-dimensional $X-Y$ models \cite{Schwabl, Maleev, Halperin, supp}.}

\AMR{While up to this point, I only considered the linear theory of interfacial active nematics, I now show that these conclusions remain unmodified by nonlinearities using standard scaling arguments. The linear anisotropy exponent implied by \eqref{ssfang} $\mu=1$ since the static structure factor of angular fluctuations diverges as $1/|q_\perp|$ in all directions of the wavevector space. Since from \eqref{angrelang} $\partial_t\theta\sim |q_\perp|\theta$ for all $\phi$, the dynamical exponent $z=1$ as well \cite{Footnote}.}
\AMR{The lowest order nonlinearity that enters \eqref{angrelang} $\sim |q_\perp|(\theta^2)_q$. The linear exponents imply that this nonlinearity is irrelevant.}
\AMR{Since all other possible nonlinearities are \emph{even less relevant}, this implies that there is no relevant nonlinearity and the exponents calculated on the basis of the linear theory are correct. Therefore, interfacial active nematics have LRO and belong to a new universality class.}

\AMR{I now discuss interactions between defects in the interfacial nematic phase. $+1/2$ defects in active nematics have a polar structure \cite{ChaiLub, deGen, Dzyaloshinskii} and are motile  \cite{Vijay, Luca2, Suraj2, Pismen1, Pismen2}. For nematics on substrates, \emph{without long-range interactions}, this leads to a defect-unbinding transition}
both at low and high noise strengths \AMR{but not at intermediate noise strengths \cite{Suraj2}}.
\AMR{A full description of defect dynamics in interfacial nematics, including their advection by the active flow leading to non-reciprocal defect interactions \cite{Vafa1, Ano_hex}, will be presented elsewhere \cite{Ano_unpub}. However, I show that the confining potential between defects is qualitatively modified by the long-range fluid interactions. }
\AMR{T}he confining potential between a \AMR{passive} $\pm1/2$ \AMR{defect} pair is $V^p_{con}\propto(K/2)\ln(|{\bf x}_\perp|/a)$ where ${\bf x}_\perp$ is the separation between the defect pair and $a$ is the size of the defect core. 
\AMR{In interfacial nematics}, the relaxation of the angular fluctuations is \AMR{controlled} by \AMR{activity-induced long-range interaction} \AMR{which vanish} only as $\sim |q_\perp|$ at small wavenumbers. 
Thus, activity \AMR{effectively} leads to an \AMR{additional elasticity} which \emph{diverges} as $1/|q_\perp|$ as $q_\perp\to 0$ or as $r_\perp$ as $r_\perp\to\infty$. Therefore, the confining potential between a widely separated $\pm 1/2$ pair \AMR{acquires an extra contribution}
$V_{con}^a\sim |{\bf x}_\perp|$ \AMR{as in passive $X-Y$ model with dipolar interaction \cite{Schwabl}}. \AMR{This directly competes against the effect of defect motility which}
within an effectively one-dimensional description, \AMR{that is essentially correct at low noise strengths,}
can be modelled as arising from a potential \AMR{$V=V^p_{con}+V^a_{con}-|v| |{\bf x}_\perp|$} where $|v|$ is the motility of the $+1/2$ defect \cite{Suraj2, Luca2}. \AMR{Since both the motility-induce unbinding and long-range-interaction-induced confinement have the same spatial character, the fate of the interfacial nematic at low noise strength depends on a detailed calculation of their relative strengths \cite{Ano_unpub}. However, this argument suggests that defects in an interfacial active nematic at low noise strengths may remain bound and the nematic phase may not be re-entrant unlike in \cite{Suraj2}.}

Till now, I considered a system in which the \AMR{number of nematogens at the interface was not constant}.
Now I consider a homogeneous system in which the concentration of active particles $c$ \AMR{at the interface or membrane} is conserved with a mean value $c_0$. 
The linearised hydrodynamic equation for the concentration field is
\begin{equation}
\label{conceq}
\partial_t c=-c_0\nabla_\perp\cdot{\bf v}+\zeta_Q\partial_x\partial_y\theta+D\nabla_\perp^2c+\xi_c
\end{equation}
where the term with the coefficient $\zeta_Q$ is the active concentration current \cite{Aditi2, RMP}, $D$ is the diffusivity and $\xi_c$ is a \AMR{spatiotemporally white} conserving noise \AMR{of strength $2\Delta^c$}. The dynamics for the angle field remains the same as \eqref{angeq} while the three-dimensional force balance equation is modified from \eqref{vel3d} to
\AMR{\begin{multline}
\label{vel3dc}
\eta\nabla^2{\bf V}=\nabla\Pi+\zeta(c_0)(\partial_y\theta\hat{x}+\partial_x\theta\hat{y})\delta(z)+{\bf f}^{p}\delta(z)+\\\zeta_1(\partial_x c\hat{x}-\partial_yc\hat{y})\delta(z)+A(c_0)\nabla_\perp c\delta(z)+\zeta_c\nabla_\perp c\delta(z)+\boldsymbol{\xi}^v
\end{multline}}
where $\zeta_1\equiv (1/2)\partial_c\zeta|_{c=c_0}$, $\zeta_c$ denotes an active isotropic stress \AMR{and $A$ is the inverse of the passive compressibility which can be tuned by changing the concentration of the active particles}. \AMR{To leading order in gradients, the concentration dynamics is controlled by $\nabla_\perp\cdot{\bf v}$}. The eigenfrequencies implied by the coupled concentration and angular dynamics, after eliminating the velocity field, are displayed in \cite{supp}.
\AMR{Importantly, for $A_r=A+\zeta_c\neq 0$, a}t least one of these eigenfrequencies has a positive imaginary part for some $\phi$ \AMR{(except in a special case discussed in \cite{supp})}, implying that the homogeneous nematic phase is destabilised, \AMR{while for $A_r=\zeta_1=0$ \AMRR{a homogeneous LRO nematic state with the angular fluctuation exponents obtained} earlier is realised  \AMRR{for $D>0$ since in this limit, the angular fluctuations are decoupled from concentration fluctuations to $\mathcal{O}(q_\perp)$}} \AMRR{(it is only possible to realise this limit when $\zeta_c<0$ since the passive compressibility $A>0$)}. \AMR{For large $A_rc_0\gg\zeta$,  i.e., high concentration of active units, the film is essentially incompressible since the dynamics is extremely sensitive to departures of $c$ from $c_0$ and $\omega_-$ has the same form as an incompressible active nematic layer in three-dimensional fluid which is generically unstable \cite{Guo1, Guo2, Sagues}, while $\omega_+$ diverges as $A+\zeta_c\to\infty$ signalling an infinitely fast relaxation of the concentration fluctuations.}
\AMR{The picture is qualitatively modified at \emph{small} $A_r c_0\ll \zeta$ which should be realised at low concentration of active units. In this case, \AMR{additionally taking $\zeta_1=0$ for simplicity}, $-i\omega_+$ equals the non-stochastic part of the R.H.S. of \eqref{angrelang} and 
	\begin{equation}
		\label{asympom}
		\lim_{A_r\to 0}\omega_-\approx-\frac{iA_rc_0|q_\perp|}{8\eta}\left[1+\frac{\lambda-4\cos 2\phi+3\lambda\cos4\phi}{-4\cos 2\phi+\lambda(3+\cos 4\phi)}\right].
\end{equation}
which is independent of $\zeta$ (this feature does not depend on the assumption of vanishing $\zeta_1$) but remains generically destabilising. \AMRR{This instability is, however, distinct from the one at large $A_r$ which depends on $\zeta$.}
When the number of active particles at the interface is constant, the nematic phase is generically destabilised and for $\zeta\lambda>0$ and $|\lambda|>1$ (such that $\omega_+$ is stabilising), the \AMRR{fastest growing wavevector} beyond the instability, \AMRR{calculated by retaining the $\mathcal{O}(q_\perp^2)$ terms in the dispersion relation,}  scales as $1/|\zeta|$ at large concentration of active units, becoming \emph{independent} of $\zeta$ at small concentration, \AMRR{at least when active corrections to the passive parameters $K$ and $D$ are small}. Alternatively, if activity is tuned, \AMRR{the growth rate} scales as $1/|\zeta|$ at small activities and saturates at a finite value at larger activities.} 

I now discuss interfacial polar phases \AMR{in which the swimmer number at the interface is conserved}. The polarisation field is described by an in-plane polar vector ${\bf p}$ which is taken to be ordered along the $\hat{x}$, ${\bf p}=p_0(\cos\theta,\sin\theta)$. The free energy, in terms of the concentration field and the angle field $\theta$ is $F=\int d{\bf r}_\perp [A\AMR{(\delta c)^2}+(K/2)(\nabla_\perp\theta)^2+\gamma\delta c\partial_y\theta]$, \AMR{where $\delta c=c-c_0$}. The equations for $c$ and $\theta$ are modified from \eqref{conceq} and \eqref{angeq} respectively:
\begin{equation}
\label{conceqp}
\partial_t c=-c_0\nabla_\perp\cdot{\bf v}-\bar{v}_cp_0\partial_x c-v_cp_0\partial_y\theta+D\nabla_\perp^2c+\xi_c,
\end{equation}
where $\bar{v}_c=\partial_c v_c|_{c=c_0}$ and 
\begin{equation}
\label{angeqp}
\partial_t\theta=-v_pp_0\partial_x\theta+\Omega_{xy}-\lambda A_{xy}-\Gamma_\theta\frac{\delta F}{\delta\theta}+\xi.
\end{equation}
The equation of motion for the three-dimensional velocity field \eqref{vel3dc} 
remains unchanged.
\AMR{In the limit of large $\AMR{A_r=}A+\zeta_c\to\infty$ \AMR{(i.e., in the effectively incompressible case)}, the \AMR{eigenfrequencies} of the coupled concentration \eqref{conceqp} and angular \eqref{angeqp}dynamics obtained after solving for the velocity field are always destabilising, just like their nematic counterparts. However, \AMRR{the instability in this case is convective and not absolute and thus} highly motile swimmers can outrun this instability \cite{Rayan} for smaller $A_r$ i.e., interfacial flocks with either large $v_p$ or large $\bar{v}_c$ ($\gg A_rc_0,\zeta$) can be stable. In either of these limits both the \AMR{eigenfrequencies} can have negative imaginary parts that vanish as $q_\perp$ implying a \emph{stable} LRO polar phase. 
}
\AMR{The \AMR{eigenfrequencies} in this case are }
\begin{equation}
\omega_+=\left[\bar{v}_cp_0\cos\phi-i\frac{c_0}{4\eta}(\AMR{A_r}+\zeta_1\cos2\phi)\right]|q_\perp|
\end{equation}
and
\begin{equation}
\omega_-=\left[{v}_pp_0\cos\phi+\frac{i\zeta}{4\eta}\left\{\cos2\phi-\frac{\lambda}{2}(1+\cos^22\phi)\right\}\right]|q_\perp|
\end{equation}
and if $\zeta_1<\AMR{A_r}$, the first \AMR{eigenfrequency} has a negative imaginary part while the second \AMR{eigenfrequency} is stabilising when $|\lambda|>1$ and $\zeta\lambda>0$ \cite{supp}. \AMR{The concentration dynamics in the LRO polar phase is anomalous -- the}
static structure factor of concentration fluctuations $\langle|c({\bf q}_\perp, t)|^2\rangle \propto 1/|q_\perp|$. The $1/|q_\perp|$ divergence implies that the R.M.S. number fluctuations $\sqrt{\langle \delta N^2\rangle}$ in a region \AMR{with $\langle N\rangle$ polar particles on average} scales as $\langle N\rangle^{3/4}$ instead of as $\langle N\rangle^{1/2}$ as it would in all equilibrium systems. This implies that the polar phase has giant number fluctuations although they are not as large as those in compressible motile fluids on substrates \cite{RMP}.
Moreover, arguments presented for the irrelevance of all nonlinearities in the apolar phase \AMR{remain} valid in this case \cite{supp} implying that the predictions of the linear theory presented here \AMR{are \emph{exact} and} should be observable in experiments .

\AMR{All the preceding discussion concerned ordering at a perfectly flat interface or membrane. In \cite{supp}, I include fluctuations of the immersed surface as well and demonstrate that these are decoupled from the planar angular dynamics. However, activity leads to an effective, anisotropic surface tension of the interface or membrane along the ordering direction, which is destabilising for extensile systems and stabilising for contractile ones. When the \emph{extensile} activity is strong enough to overcome the passive surface tension, the flat state itself becomes unstable \cite{Ano_und, supp} to buckling and probably assumes an undulated form \cite{Ano_und}. While the theory presented in this paper is for a flat surface, and therefore, for contractile systems and extensile systems with not too large a value of activity, it is possible that the in-plane orientational dynamics described here remains valid even on a corrugated interface or membrane since the angular dynamics is decoupled from shape fluctuations at linear order.}

\AMR{I close with a discussion of experimental systems in which the theory presented here may be tested and interfacial uniaxial order may be observed. Motor-microtubule layers immersed in fluids have been one of the mainstays of studying active pattern formation \cite{Dogic1, Dogic2, Dogic3, Sagues, Ano_und, Ramin_und}. Here, the number of capped microtubule filaments at the interface is essentially constant and nematic state should be generically unstable. The prediction that the \AMRR{fastest-growing mode} beyond the instability becomes effectively independent of activity at high activities or at small filament concentration can be checked either by increasing kinesin concentration or decreasing microtubule concentration in this system. More radically, an LRO nematic phase may be realised in this setup by allowing the microtubule filaments to associate and dissociate at the interface and the monomers to diffuse in the fluid. Interfacial active nematics can also be created by impregnating passive interfacial nematics with bacteria \cite{LLC1, LLC2}. Beyond such biomimetic systems, interfacial or membrane-associated ordering may be widely present in biological systems ranging from bacteria to cellular cortex. In this context, polar ordering of short actomyosin filaments associated with the cell membrane immersed in a bulk, cytoskeletal, cortical fluid may be especially important since they have been argued to be involved in the active transport leading to nanoclustering of cell-surface molecules \cite{jitu1, jitu2, Madan1, Madan2, Madan3}.  
	Finally, while the paper considered order at an immersed surface, since the key properties required for the reported LRO uniaxial phases are the lack of surface incompressibility and long-range fluid interactions, qualitatively similar results should also hold for uniaxial ordering on the interface of a semi-infinite fluid medium \cite{Niladri1, Niladri2}.}
\begin{acknowledgments}
I thank Sriram Ramaswamy and Raphael Voituriez for insightful comments and discussions. I also thank Sriram Ramaswamy for a careful reading of the manuscript and crucial suggestions.
\end{acknowledgments}


\begin{thebibliography}{}
\bibitem{Footnote1} \AMR{While it has been shown in \cite{Dogic_incmp} that the microtubule filaments at the interface are essentially incompressible, this is likely an artefact of their concentration -- or that of surfactants -- in that experiment; a less dense aggregate of microtubule filaments should be effectively compressible at large scales.}
\bibitem{Footnote} The linear exponents can also be obtained by demanding that the damping term $\propto |q_\perp|$ and the noise strength does not change under rescaling for all directions of the wavevector space. This yields the same linear exponents.
\end{thebibliography}
\end{document}